\documentclass[aps,prl,twocolumn,superscriptaddress]{revtex4-1}

\usepackage{graphicx}
\usepackage{amsmath}

\begin{document}


\title{Phase separation of self-propelled ballistic particles}


\author{Isaac R. Bruss}
\affiliation{Chemical Engineering, University of Michigan, Ann Arbor, MI 48109, USA}
\author{Sharon C. Glotzer}
\email[]{sglotzer@umich.edu}
\affiliation{Chemical Engineering, University of Michigan, Ann Arbor, MI 48109, USA}
\affiliation{Materials Science \& Engineering, University of Michigan, Ann Arbor, MI 48109, USA}
\affiliation{Biointerfaces Institute, University of Michigan, Ann Arbor, MI 48109, USA}


\date{\today}

\begin{abstract}
Self-propelled particles phase separate into coexisting dense and dilute regions above a critical density. The statistical nature of their stochastic motion lends itself to various theories that predict the onset of phase separation. However, these theories are ill equipped to describe such behavior when noise become negligible. To overcome this limitation, we present a predictive model that relies on two density-dependent timescales: $\tau_F$, the mean time particles spend between collisions; and $\tau_C$, the mean lifetime of a collision. We show that only when $\tau_F < \tau_C$ do collisions last long enough to develop a growing cluster and initiate phase separation. Using both analytical calculations and active particle simulations, we measure these timescales and determine the critical density for phase separation in both 2D and 3D.

\end{abstract}


\maketitle

Statistical physics extracts order from randomness by describing the average behavior of noisy systems. Such noise is prevalent in active matter \cite{Marchetti2013a, Ramaswamy2010}, where individual particles generate their own motion by consuming energy from their environment, whether it be from chemical reactions \cite{Palacci2013, Palacci2015a, Volpe2011}, vibrations \cite{Narayan2007, Deseigne2010}, light \cite{Jiang2010}, or magnetic fields \cite{Snezhko2011, Snezhko2009a}. However, such active systems fall outside the realm of equilibrium statistical physics because, although they may be at steady state, they are inherently non-equilibrium and do not obey detailed balance \cite{Fodor2016a}.

\begin{figure}[b]
\centering
\includegraphics[width=0.5\textwidth]{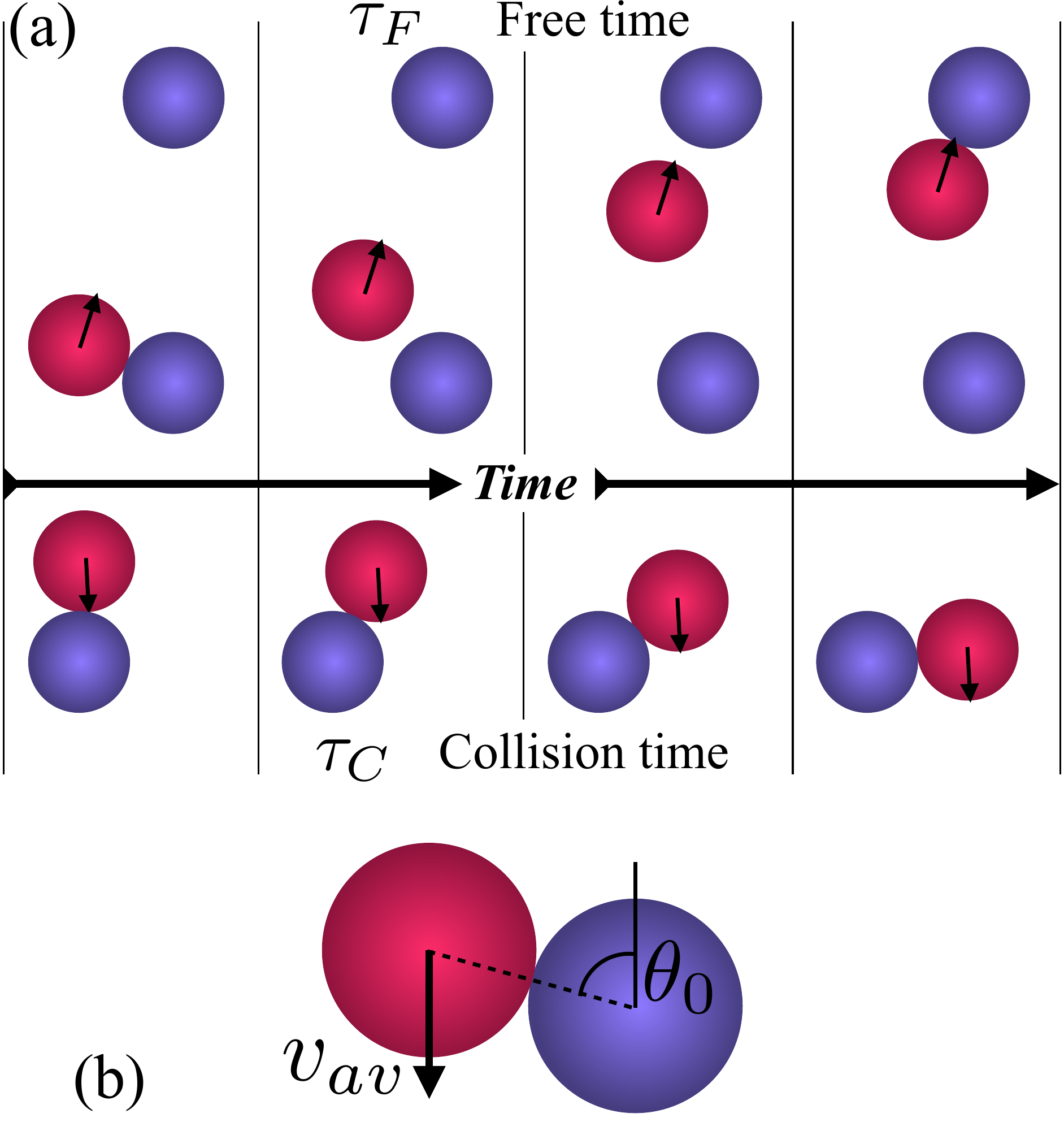}
\caption{(a) A schematic representation of two key timescales involved in the triggering of phase separation for active self-propelled particles. For a single active particle (red), the free time $\tau_F$, considers the time a particle spends between collision events, while the collision time $\tau_C$, considers the lifetime of a collision. (b) A schematic representation of the incident angle of collision $\theta_0$, for eqn (\ref{eqn:tauCIntegrand}).}
\label{fig:schematics}
\end{figure}

Despite these challenges, emergent behaviors are still observed in active matter systems, such as their ability to phase separate. In particular, active Brownian particles (ABP), an idealized system of self-propelled hard particles, each with a rotationally diffusing direction of motion \cite{Cates2013a}, have been found to phase separate into dense and dilute phases \cite{Fily2012, Redner2013a}. This robust behavior has also been found for both two and three dimensions \cite{Stenhammar2014a}, as well as for rod-like \cite{Wensink2012a, Abkenar2013, Mishra2014} and shaped particles \cite{Wensink2014, Ilse2016a}, attractive particles \cite{Redner2013, Prymidis2015c}, contact-triggered active particles \cite{agrawal2017tunable}, and mixtures of ABP with passive particles \cite{Stenhammar2015, Wysocki}.

Two prevailing theories describe motility-induced phase separation (MIPS) for ABP. One is a kinetic theory that balances the density-dependent inward flux of particles into the dense phase, with a diffusion-dependent outward flux \cite{Redner2013a, Redner2016}. The other is a continuum mean field theory that attributes phase separation to the reduction in a particle's effective speed by an increasing local density \cite{Stenhammar2013, Cates2015a, Nardini2017}. Critical to both of their forumaltions is the noise involved in the rotational diffusion of the particles. Given a rotational diffusion constant of $D_R$, the persistence length, $\ell_P$, of an active particle moving with a velocity $v_0$ is $v_0/D_R$. For the kinetic theory, the rotational diffusion regulates the rate at which boundary particles become unblocked and escape the dense phase. This balance leads to the incorrect prediction in the limit of $\ell_P \to \infty$, of a critical density for phase separation of $\phi_{crit} = 0$ \cite{Redner2013a}. However, previous simulations at or approaching this ballistic regime show a nonzero critical density for MIPS (via spinodal decomposition) of $0.25 \gtrsim \phi_{crit} \gtrsim 0.35$ in 2D \cite{Cates2015a, Fily2014, Solon2016, Levis2017}. The range of reported densities is likely due to a small system size, with higher $\phi_{crit}$ being measured for numbers of particles less than 10,000 \cite{Bruss2017}. Also, these results are for simulations where translational diffusion is zero, which approximates the behavior of run-and-tumble bacteria \cite{Cates2013a, Tailleur2008}. When translational diffusion is proportional to $D_R$, $\phi_{crit} \approx 0.4$ \cite{Fily2012, Stenhammar2013}. Finally, like the kinetic theory, the continuum theory of MIPS is only meaningful when noise is present for finite $\ell_P$, because the effective pressure diverges when $D_R = 0$ \cite{Cates2015a, Solon2016}. And while results for large $\ell_P$ can still be extracted, numerical fitting parameters are required \cite{Solon2016}.

Although these two theories explain MIPS in their respective regimes, and can be used to extract other important properties such as cluster fraction and growth rates, there exists a gap in the understanding of MIPS when rotational noise is absent (or, at minimum, when $\ell_P$ is much greater than a particle's diameter). Examples of such ballistic-regime active matter systems are straight-swimming bacteria that do not tumble due to mutations \cite{Aswad1974} or hydrodynamic suppression \cite{Molaei2014}, and vibrated polar disks \cite{Deseigne2010}. Here we present a new model of MIPS that is independent of rotational noise, and sets a lower bound on the critical density required for phase separation. Although a system with zero noise is not strictly ergodic for a finite number of particles, we argue that this detail is irrelevant for phase separation because the initial stages of spinodal decomposition happens on a finite length scale where particle collisions can be considered ergodic.

We demonstrate that the MIPS of ABP systems is dependent on the relation between two density-dependent timescales (shown schematically in Fig.~\ref{fig:schematics}(a): first, the mean free time, $\tau_F$, defined as the average span between collisions, which at low densities is synonymous with the mean free path divided by the particle's speed; and second, the mean collision time, $\tau_C$, defined as the average lifetime of a two-particle collision. We propose that at the density where $\tau_C \geq \tau_F$, which we call the \emph{congestion density}, $\phi_{con}$, particles experience a rate of collisions, $\tau_F^{-1}$, that outpaces their rate of separating from a collision, $\tau_C^{-1}$. This imbalance results in a positive feedback effect of particle congestion, which drives the formation of ABP clusters. Thus $\phi_{con}$ sets a strict lower bound on the value of $\phi_{crit}$ required for spinodal phase separation.

Both of these timescales can be calculated for ballistic ABP, which we will later compare to simulation results. First, the mean free time, $\tau_F$, can be determined from collision theory \cite{Pilling1995}. Beginning with 2D, we know that $N$ particles, traveling with a speed $v_0$ at low densities, will each collide once on average by the time they have moved through an area equal to the averaged area available to them. This area is simply the total area divided by $N$, or $\frac{\pi \sigma^2}{4 \phi}$, where $\sigma$ is the particles' diameter, and $\phi$ is the area fraction of particles. The area swept out by an ABP is $2 \sigma v_{av} \tau_F$, where $v_{av} = \frac{1}{\pi} \int^{\pi}_0  \sqrt{2-2\cos \theta} d\theta = 4v_0/\pi$ is the relative velocity of an active particle compared to all other particles, averaged over all possible relative angles $\theta$ \cite{Pilling1995}. Likewise for 3D, the {\em available volume} is $\frac{\pi \sigma^3}{6 \phi}$, while the {\em swept volume} is $\pi\sigma^2v_{av}\tau_F$, with $v_{av} = \frac{1}{4\pi} \int^{2\pi}_0 \int^{\pi}_0 \sin \theta \sqrt{2-2\cos \theta} d\theta d\psi = 4v_0/3$. Setting these two areas (or volumes in 3D) equal to each other, the resulting mean free time for a particle to wait between collisions is
\begin{align}
\tau_F = \frac{\pi^2 \sigma}{32 v_0 \phi} ~~~~~~~~ (2D) \nonumber \\
\tau_F = \frac{\sigma}{8 v_0 \phi}. ~~~~~~~~ (3D)
\label{eqn:tauF}
\end{align}

\begin{figure*}[t]
\centering
\includegraphics[width=1\textwidth]{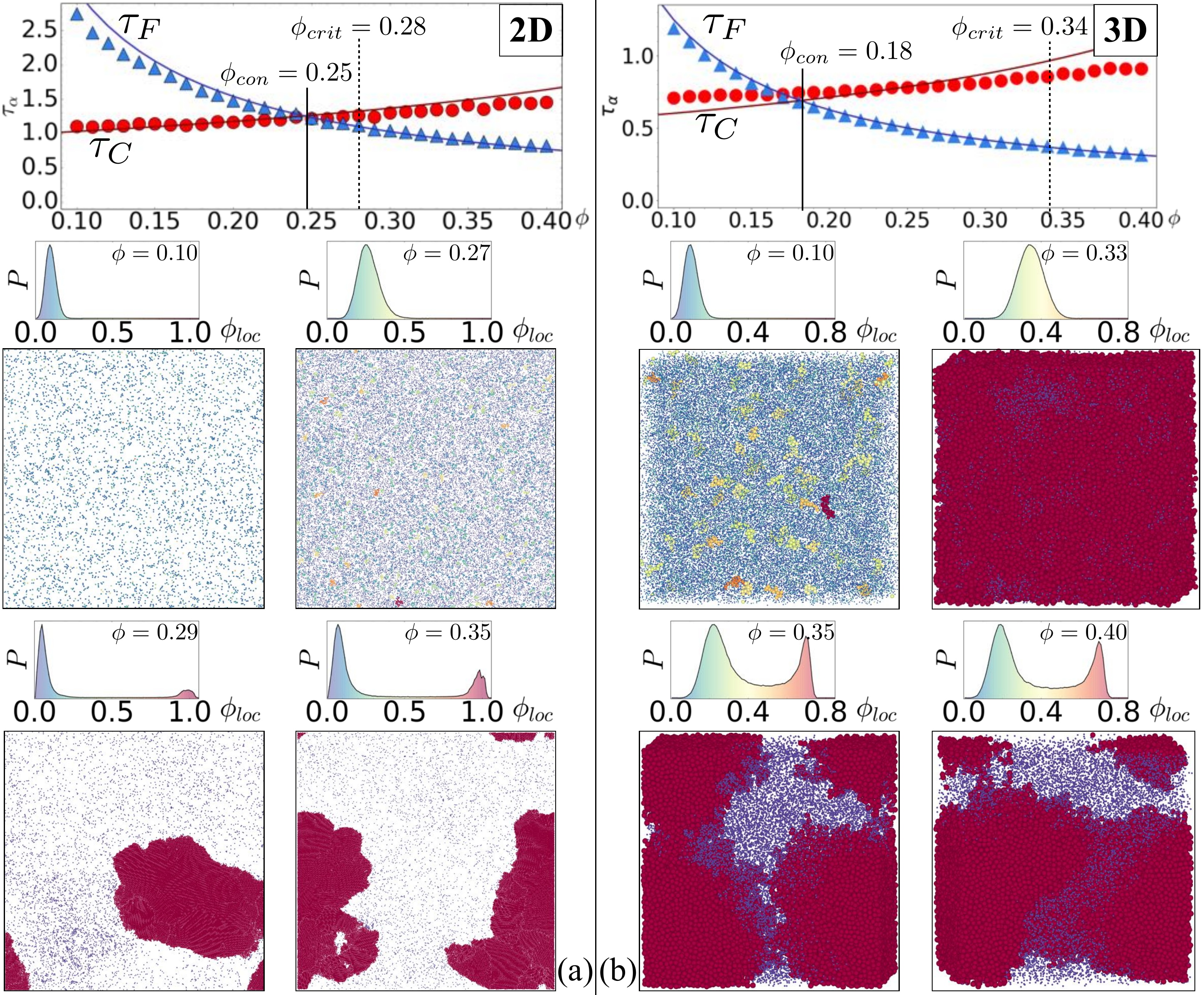}
\caption{The mean free time $\tau_F$ (blue triangles), and mean collision time $\tau_C$ (red circles), as a function of area fraction $\phi$, in both 2D (a) and 3D (b). Solid colored lines represent the predicted values calculated with eqns (\ref{eqn:tauF}\&\ref{eqn:tauC}). The congestion density, being the minimum density required for phase separation is the point where $\tau_C > \tau_F$, is $\phi_{con} > 0.25$ for 2D, and $\phi_{con} > 0.18$ for 3D (marked with vertical solid lines). The critical density for phase separation found using simulations, is $\phi_{crit}$ (marked with the vertical dashed lines). Visualizations of the results at various densities are shown with particles colored by their cluster size. Local density histograms are also provided (using the same color scale), showing a transition from unimodal to bimodal distributions upon phase separation. For the 3D visualization, particles not belonging to a large cluster are reduced in diameter for easier viewing.}
\label{fig:timescales}
\end{figure*}

To solve for the mean collision time $\tau_C$, we consider on average how long two frictionless ABP remain in contact when colliding. We frame a collision event such that one particle is held fixed and the other is moving downwards with a velocity $v_{av}$, and colliding at an incident angle $\theta_0$, as shown in Fig.~\ref{fig:schematics}(b). In this instance, the incident angle-dependent collision time $\tau_{inc}$, is simply the time it takes the ABP to move around its fixed neighbor and sweep out the angle between $\theta_0$ and $\pi/2$
\begin{equation}
\tau_{inc}(\theta_0) = \int_{\theta_0}^{\pi/2} \frac{\sigma d\theta}{(1-\phi^*)v_{av} \sin \theta}
\label{eqn:tauCIntegrand}
\end{equation}
The correction to the velocity, $(1-\phi^*)$, linearly interpolates between the ideal unhindered movement of two isolated particles, and the fully arrested state of particles found at a limiting density $\phi^*$. Such a phenomenological factor has been measured directly in systems of active Brownian particles, and is predicted from various kinetic theories \cite{Stenhammar2013, Fily2014, Stenhammar2014a, Cates2015a}. Our application of this first-order correction to the collision velocity can be attributed to multi-body collisions that act to hinder particle movement, and gives $\tau_C$ its necessary density dependence. Although this linear relationship becomes inaccurate at higher densities, simulations show that it holds for $\phi \lesssim 0.5$ \cite{Cates2015a}, which includes both the density ranges where $\tau_C \approx \tau_F$, and where phase separation occurs. Averaging over all possible incident angles $\theta_0$, in both 2D and 3D yields
\begin{align}
\tau_C = \frac{G}{1-\phi/\phi^*_{2D}} ~~~~~~~~ (2D) \nonumber \\
\tau_C = \frac{\ln(8)}{4(1-\phi/\phi^*_{3D})}, ~~~~~~~~ (3D)
\label{eqn:tauC}
\end{align}
where $G \approx 0.916$ is Catalan's constant, and we assume movement becomes fully arrested at the area fraction of hexagonally packed disks in 2D, $\phi^*_{2D} \approx 0.907$, and the volume fraction of hexagonally close packed spheres in 3D, $\phi^*_{3D} \approx 0.740$ \cite{weaire2008pursuit}. (Previous studies estimate these densities using numerical fitting methods, which result in larger values of $\phi^*_{2D} \approx \phi^*_{3D} = 0.95$ \cite{Stenhammar2013, Stenhammar2014a}.) Both $\tau_F$ and $\tau_C$ are plotted with respect to the average area fraction (or volume fraction in 3D), $\phi$, as solid lines in Fig.~\ref{fig:timescales}. Adapting this calculation to various experimental active systems may require corrections that account for hydrodynamic effects, friction, and multi-body collisions at larger densities; however, this simple first-order approximation is sufficient for an initial estimate of $\phi_{con}$. Solving for the critical density for MIPS when $\tau_C = \tau_F$, yields $\phi_{con} = 0.25$ for 2D, and $\phi_{con} = 0.18$ for 3D. We predict that for densities above $\phi_{con}$, ABP will experience on average a higher rate of new collisions compared to terminating collisions, thus resulting in a pileup of ABP and phase separation.

To verify these values and to explicitly determine both $\phi_{con}$ and $\phi_{crit}$, we perform active particle dynamics simulations that obey the Brownian equations of motion
\begin{equation}
\gamma \dot{\bf r}_i = {\bf F}_i^{SP} + \sum_j {\bf F}_{ij}^{Ex}, ~~~~~~~~~~~~\dot{\theta}_i = \sqrt{2 D_R} \eta_i,
\label{eq:Brown}
\end{equation}
where $\gamma$ is the drag coefficient and $\eta_i$ is Gaussian white noise, with $\langle \eta_i(t) \rangle = 0$, and $\langle \eta_i(t) \eta_j(t') \rangle = \delta_{ij} \delta(t-t')$ \cite{Henkes2011}. Each ABP experiences a self-propulsion force ${\bf F}_i^{SP} = v_0 \hat{n}_i = v_0 (\cos \theta_i, \sin \theta_i)$ that undergoes rotational diffusion scaled by $D_R$. Nearby ABP interact via a frictionless excluded-volume repulsive force ${\bf F}_{ij}^{Ex}$, given by the steeply repulsive Weeks-Chandler-Andersen potential (i.e.\ just the repulsive portion of the Lennard-Jones potential), with the particle diameter $\sigma$, defined as the distance where ${\bf F}_{ij}^{Ex} = 0$ \cite{JohnD.WeeksDavidChandler1971}. The persistence length of the self-propulsion path is held constant at $\ell_P = 1{,}000 \sigma$. This value is chosen because $\ell_P/v_0$ is several orders of magnitude greater than the timescales of $\tau_C$ and $\tau_F$, and therefore it should not effect the onset of MIPS. To corroborate this assumption, previous studies have already reported a $\phi_{crit}$ that is independent of $\ell_P$ for $\ell_P \gtrsim 100 \sigma$ \cite{Cates2015a, Fily2014, Solon2016}. Time is measured in units of $\tau = \sigma/v_0$. The area fraction covered by $N$ particles is $\phi = \frac{N \pi\sigma^2}{4 A_{total}}$ in 2D, and $\phi = \frac{N \pi \sigma^3}{6 V_{total}}$ in 3D. All simulations were performed using the particle simulation toolkit, HOOMD-blue (version 2.1) \cite{Anderson2008}, with a step size of $10^{-4} \tau$.

Simulations used to measure $\tau_F$ and $\tau_C$ were performed for $N=2{,}000$ particles in both 2D and 3D. Measurements were taken before steady state was reached by averaging over only the first $30 \tau$ of the simulations. This choice permits a close comparison to the ideal gas regime of the analytical results of eqns (\ref{eqn:tauF}\&\ref{eqn:tauC}). Additionally, a system comprised of only 2,000 particles still guarantees that ABP have sampled only a fraction of the periodic simulation box, negating the typical requirement of $N\gtrsim10{,}000$ necessary to ignore finite number effects that have been observed in previous studies \cite{Bruss2017}. The initial state was generated by placing particles at random locations, and then relaxing their positions via a repulsive spring force to remove all particle overlaps. The direction of the self-propulsion force for each particle, $\hat{n}_i$, was also assigned randomly. This initial state approximates equilibrated passive particles that then have their self-propulsion activated. The specific procedure to generate the initial random state was found to have little effect on the measurements of $\tau_C$ and $\tau_F$.

\begin{figure}[t]
\centering
\includegraphics[width=0.5\textwidth]{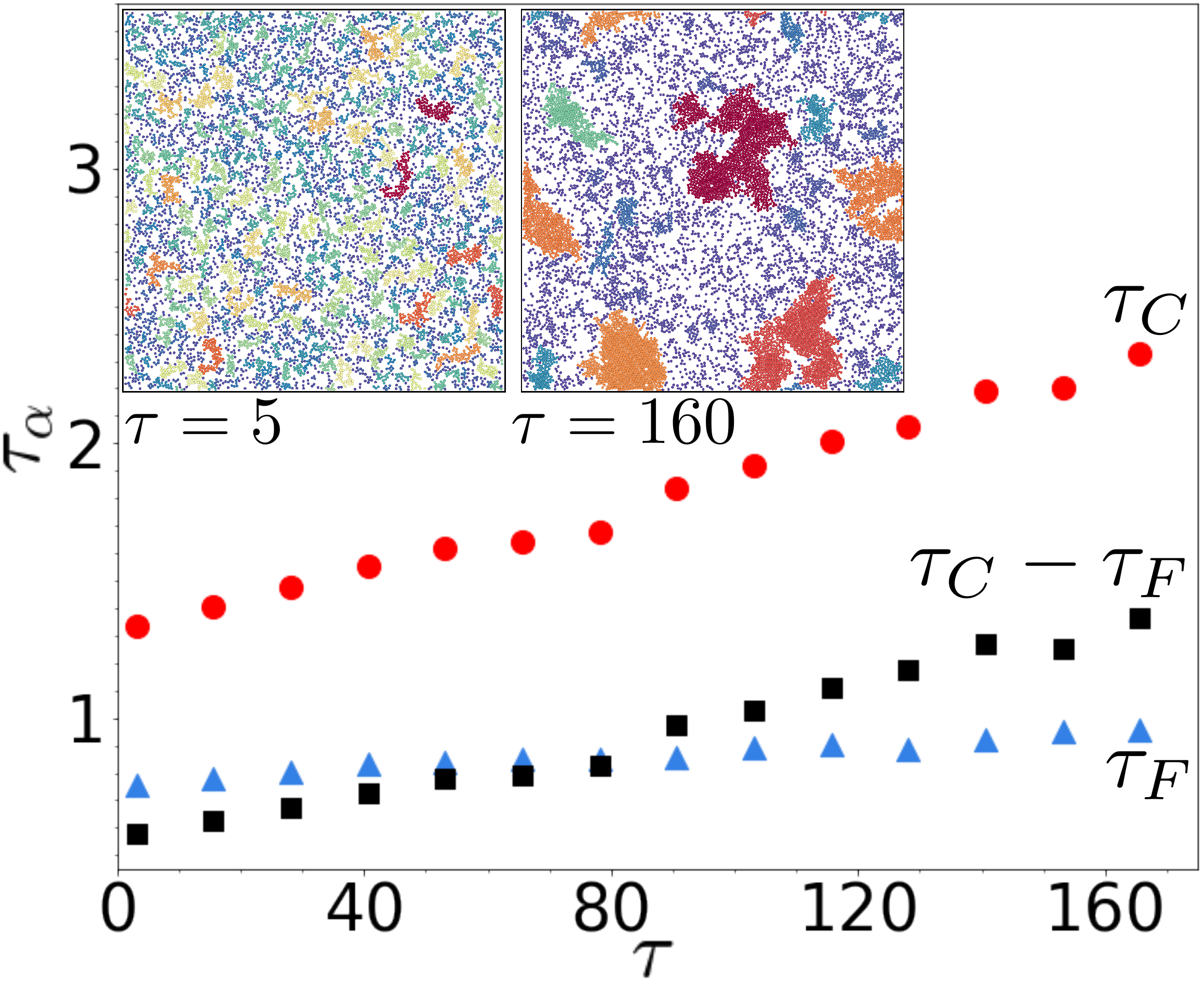}
\caption{Timescales $\tau_C$ (solid red line), $\tau_F$ (solid blue line), and $\tau_C - \tau_F$ (dashed black line), as a function of time $\tau$ in 2D. The density is $\phi = 0.4$, well above the spinodal of phase separation at $\phi_{crit} = 0.28$. Insets show snapshots at two different times, with particles colored by their cluster size. Each data point is averaged over at least 10,000 measurements.}
\label{fig:longtime}
\end{figure}

$\tau_F$ is calculated as the mean duration that particles spend completely free of contact (defined as ${\bf F}_{ij}^{Ex} = 0$), while $\tau_C$ is calculated as the mean duration of any two-particle contacts. Evidently, the accessible timescales of the initial state are sufficient to determine the final steady-state behavior. The short $30 \tau$ period of time is sufficient to measure the desired timescales well before phase separation. This is true even in the worst case scenario when the density is well above $\phi_{crit}$, where clusters quickly begin to nucleate and grow, thus greatly changing $\tau_C$ and $\tau_F$. To demonstrate this point, Fig.~\ref{fig:longtime} shows steadily increasing collision and free timescales for $\phi = 0.4$. By $30 \tau$, both timescales have increase by less than 10\%. The increase of $\tau_C$ with time is explained by ABP becoming trapped in ever-growing clusters, which can be seen in the inset visualizations at $\tau=5$ versus $\tau=160$. Meanwhile, $\tau_F$ also increases with time because there are fewer particles in the dilute phase, thus increasing the mean free path of active particles that are not yet trapped in clusters. Most importantly however, the difference between the two timescales, $\tau_C - \tau_F$, also grows with time, which acts to further increase the drive for phase separation. Eventually, the phase separated steady state is reached (after $\sim10^5\tau$), resulting in plateaued values of $\tau_C$ and $\tau_F$ with time. In other words, a system undergoing phase separation enters a positive feedback loop where clustering decelerates particles, which locally increases the lifetime of collisions $\tau_C$, which leads to growing clusters, which decelerates more particles, and so on. Alternatively, for densities below $\phi_{crit}$, clusters remain relatively small and short-lived, effecting little change in $\tau_C$ and $\tau_F$ beyond what is measured over the first $30 \tau$. Overall, this behavior confirms that the density $\phi_{con}$ at which $\tau_C > \tau_F$, is an necessary lower bound for $\phi_{crit}$. In the end, we are able to obtain accurate statistics using only the first $30\tau$ of an initially random system.

Measurements for $\tau_C$ and $\tau_F$ under these simulation parameters for both 2D and 3D are shown in Fig.~\ref{fig:timescales}. The results fit well with the analytical solutions of eqns (\ref{eqn:tauF}\&\ref{eqn:tauC}), even up to $\phi=0.4$. This density regime is well beyond where the assumptions of low density are expected to hold; namely that only uninterrupted two-body collisions are present (assumed when approximating $\tau_C$), and, more surprisingly, that all collisions can be ignored and our system behaves as an ideal gas (assumed when approximating $\tau_F$).

Now that $\tau_C$ and $\tau_F$, are established, we show that  the inequality, $\tau_C > \tau_F$, is indeed a minimal requirement for MIPS of ABP. To determine the $\phi_{crit}$ at which this happens, we perform full simulations of $N=50{,}000$ particles for $4\times10^7 \tau$. This ensures measurements are made after reaching the final steady state behavior, and that finite number (and finite persistence length) effects can be ignored. The measurement for $\phi_{crit}$ are marked as the vertical dashed lines in Fig.~\ref{fig:timescales}. Phase separation is observed at $\phi_{crit} \approx 0.28$ for 2D, and $\phi_{crit} \approx 0.34$ for 3D, determined by the development of a bimodal distribution in the local density histogram \cite{Redner2013a}. Example histograms, along with visualizations of the systems at four values of $\phi$, both near and far from the transition, are provided in Fig.~\ref{fig:timescales}. Both 2D and 3D measurements of $\phi_{crit}$ are above the minimum values of $\phi_{con}$ predicted from both eqns (\ref{eqn:tauF}\&\ref{eqn:tauC}).

\begin{figure}[h]
\centering
\includegraphics[width=0.5\textwidth]{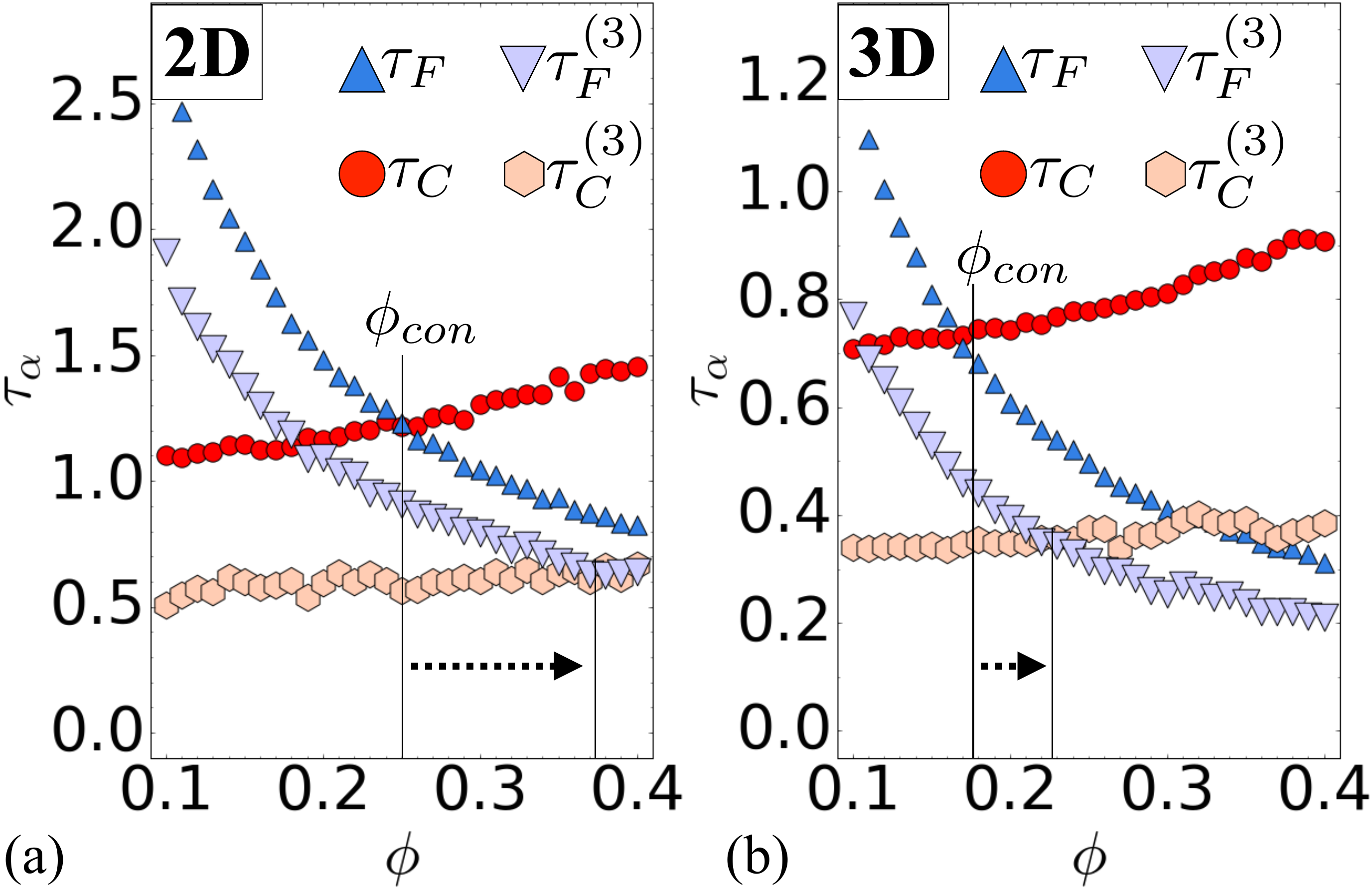}
\caption{Average timescale measurements from simulations for clusters consisting of $n=3$ particles in both (a) 2D, and (b) 3D. The average individual timescales for all particles from Fig.~\ref{fig:timescales} are also shown for comparison. Larger clusters tend to have a crossover of $\tau_C^{(n)} = \tau_F^{(n)}$ at a density higher than $\tau_C = \tau_F$ (marked with vertical lines).}
\label{fig:clusterTimes}
\end{figure}

The reason $\phi_{crit} \geq \phi_{con}$, is because $\phi_{con}$ sets the minimum density required for 2-particle clusters to, on average, live long enough to grow. However, this behavior is not sufficient to attain full phase separation at $\phi_{crit}$. This is because $\phi_{con}$ assumes only ideal two-particle collisions at low density; but in reality, the formation of the dense phase requires many multi-particle collisions. Therefore, one must consider the lifetimes of these multi-particle clusters to determine if they themselves grow into a larger phase-separated cluster, or disperse back into the dilute phase. To quantify this behavior, if $\tau_C$ and $\tau_F$ measure the average timescales experienced by a single particle, we define a new similar quantities for a dense cluster consisting of $n$ particles, where the corresponding rates are not simple linear functions of $n$. Specifically, one can argue that the average time between particles leaving a cluster of size $n$ is $\tau_C^{(n)} \gtrsim \tau_C/n$, because multi-particle clusters can cooperatively act to stabilize a collision and increase its lifetime; and similarly, the average time between particles adding to a cluster is $\tau_F^{(n)} \gtrsim \tau_F/n$, because the collision cross section of a cluster is on average less than $n 2 \sigma$ (or $n \pi \sigma^2$ in 3D). Values for $\tau_C^{(3)} = \tau_F^{(3)}$, measured in simulations for clusters of size $n=3$ are shown in Fig.~\ref{fig:clusterTimes}, which verify the above inequalities. The end result is that $\phi_{con}$, which is set by when $\tau_C = \tau_F$, is not sufficient alone to determine $\phi_{crit}$; nevertheless, this timescale analysis does yield an accurate lower bound density for MIPS.

Furthermore, these multi-particle collisions explain why the inequality of $\phi_{crit} \geq \phi_{con}$ is greater in 3D than 2D, as is obvious in Fig.~\ref{fig:timescales}. This behavior can be largely attributed to the fact that the extra third dimension provides an additional direction for particles to slide past each other. Whereas in 2D, two particles in contact occupy $1/6^{th}$ of their available solid angle; in 3D, this decreases to $<1/13^{th}$ \cite{weaire2008pursuit}. Therefore, 3D particles are much less likely to pileup into multi-particle clusters with many contacts, and consequently have a much smaller $\tau_C^{(n)}$ than 2D particles, for all values of $n$. This results in 3D multi-particle clusters that are too short-lived to induce phase separation, which is only overcome at densities well above $\phi_{con}$.

In conclusion, by comparing the timescales associated with collisions of ABP, we have determined a lower bound on the critical density, $\phi_{crit}$, required for MIPS. Our findings provide microscopic insight into the mechanism of early-stage MIPS, and supplement the existing kinetic and continuum-based theories by describing the behavior of ABP in the ballistic regime. This theory uniquely describes the ballistic regime of $\ell_P/\sigma \to \infty$, where a strict notion of global ergodicity is lost; however, ergodicity persists on the small size and time scales relevant to early stage phase separation, which allows for the determination of a system's average behavior. Importantly, we determine that, at minimum, phase separation requires a density high enough to achieve congestion, where the mean lifetime of a collision, $\tau_C$, is larger than the mean time between collisions, $\tau_F$. Surprisingly, a simple ideal gas regime calculation---resulting in eqns (\ref{eqn:tauF}\&\ref{eqn:tauC})---is sufficient to calculate $\phi_{con}$, which in turn sets a lower bound for $\phi_{crit}$.

From this foundation, one can now account for the rotational noise of ABP and allow $\tau_C$ and $\tau_F$ to be functions of $\ell_P$. Interestingly, as long as $\ell_P \gtrsim \sigma$, this results in $\tau_F(\ell_P)$ being equal to the $\tau_F$ from eqn (\ref{eqn:tauF}), because the path of an active particle still remains relatively straight in this regime. Similarly when $\ell_P \gtrsim \sigma$, $\tau_C(\ell_P)$ remains relatively unchanged from eqn (\ref{eqn:tauC}), because the timescale of rotational diffusion, $D_R^{-1}$, is insignificant when larger than the average lifetime of a two-particle collision. A full analysis, both numerical and analytical, concludes with a predicted $\phi_{con}$ that increases only when $\ell_P \lesssim \sigma$. This outcome clearly fails to account for the observed increase in $\phi_{crit}(\ell_P)$, for as early as $\ell_P \lesssim 100 \sigma$ \cite{Cates2015a, Fily2014, Solon2016}. In this regime our claims of a lower bound on $\phi_{crit}$ set by $\tau_C(\ell_P) > \tau_F(\ell_P)$ still hold true; however, there emerges a new $\ell_P$-dependent length scale that accounts for the rate of particles escaping a cluster into the dilute phase. These events act to further regulate large-scale phase separation by increasing $\phi_{crit}$ with decreasing $\ell_P$ \cite{Redner2013a, Redner2016}.

Beyond rotational noise, there are a variety of additional considerations that can affect $\phi_{con}$, and therefore $\phi_{crit}$. These include: ({\em i}) the role of particle shape, which may act to increase $\tau_C$, and thus $\phi_{crit}$, if facets are present \cite{Wensink2014, Ilse2016a}, or alternatively to decrease $\phi_{crit}$ by increasing $\tau_F$ in the case of active rods that have a tendency to swarm \cite{Wensink2012a, Abkenar2013, Mishra2014}; ({\em ii}) the role of passive particles mixed with ABP, which increases $\phi_{crit}$ by perhaps increasing the effective $\tau_F$ \cite{Stenhammar2015, Wysocki}; ({\em iii}) the influence of fixed barriers, which decrease $\phi_{crit}$, conceivably by trapping particles and greatly increasing $\tau_C$ \cite{Kaiser2012, Reichhardt2014}; ({\em iv}) the role of inter-particle attraction, which would lower $\phi_{crit}$ through a simultaneous increase of $\tau_C$ and decrease of $\tau_F$ \cite{Redner2013}; ({\em v}) the role of particle elasticity, which could see an increase in $\phi_{crit}$ for ``softer'' particles through a lowering of $\tau_C$, as suggested by an increased $\phi_{crit} = 0.46$ found for ballistic active particles with soft harmonic repulsion \cite{Reichhardt2014}; ({\em vi}) the role of particle ``eccentricity'' (i.e.\ orbiting ABP with an off-centered self-propulsion force), which can lose their ability to phase separate, possibly because rotating particles can easily slide past each other and lower their $\tau_C$ \cite{Ma2017}; ({\em vii}) the complex role of hydrodynamics, which may result in either enhanced or diminished phase separation depending on the if the particles are ``pushers'' versus ``pullers'', and/or if they are fully 3D versus confined quasi-2D \cite{Zottl2014, Matas-navarro2014, Blaschke2016, Oyama2016}; ({\em viii}) or the role of non-negligible particle inertia \cite{Golestanian2009}. These examples demonstrate that the multitude of active matter systems beyond the {\em vanilla} ABP model, possess a rich behavior that could possibly be understood in terms of congestion on the microscopic scale. However, a full quantitative exploration of $\tau_C$ and $\tau_F$ in each variant is left as an open question.

\begin{acknowledgments}
We thank Shannon Moran and Mayank Agrawal for helpful feedback, also Michael Hagan and Michael Cates for insightful discussions. This work was supported as part of the Center for Bio-Inspired Energy Science, an Energy Frontier Research Center funded by the U.S. Department of Energy, Office of Science, Basic Energy Sciences under award DE-SC0000989. Computational resources and services were supported by Advanced Research Computing at the University of Michigan, Ann Arbor.
\end{acknowledgments}

\bibliography{apstemplate}

\end{document}